\documentstyle[12pt,a4]{report}
\def\be{\nopagebreak[3]\begin{equation}}
\def\ee{\end{equation}}
\def\ba{\begin{array}}
\def\ea{\end{array}}

\def\ni{\noindent}
\newcommand{\scr}[1]{{\scriptstyle #1}}
\newcommand{\scs}[1]{{\scriptscriptstyle #1}}
\newcommand{\R}{{\scriptscriptstyle R}}
\newcommand{\T}{{\scriptscriptstyle T}}
\begin{document}
\bibliographystyle{unsrt}
\begin{titlepage}
\begin{flushright}
 NBI-HE-92-66\\
October, 1992
\end{flushright}
\vspace*{2pc}
\begin{center}
{\Huge \bf
q-deformed lattice gauge theory\\
and 3-manifold invariants}
\end{center}
\vspace{2pc}
\begin{center}
 {\Large D.V. Boulatov}\\
\vspace{1pc}
{\em The Niels Bohr Institute\\
University of Copenhagen\\
Blegdamsvej 17, DK-2100 Copenhagen \O \\
Denmark}
\vspace{3pc}

{\large\bf Abstract}
\end{center}

The notion of $q$-deformed lattice gauge theory is introduced.
If the deformation parameter is a root of unity, the weak
coupling limit of a 3-$d$ partition function gives a topological
invariant for a corresponding 3-manifold. It enables us to define the
generalized Turaev-Viro invariant for cell complexes.
It is shown that this invariant is determined by an action of a
fundamental group on a universal covering of a complex.
A connection with invariants of framed links in a manifold is also
explored. A model giving a generating function of all simplicial
complexes weighted with the invariant is investigated.

\end{titlepage}

\vspace*{4pc}

\stepcounter{chapter}
{\Large\bf Introduction}

\vspace{2pc}

The last few years some complex of ideas and methods has been formed
linking together 3-dimensional topology, topological quantum field
theories and 3-d quantum gravity. The key-stone here is quantum groups,
which have provided us with a number of formulas miraculously appearing in
different at first sight contexts. Developed as the mathematization
of algebraic structures appearing in exactly soluble 2-dimensional
models, \cite{Drin}
quantum groups are now considered as one of fundamental concepts
in modern mathematical physics. Inherited many properties of ``classical''
group theory, they non-trivially generalize the notion of symmetry. Their
application has led to recent progress in knot theory and allowed for
the construction of various 3-manifold invariants. After the pioneering
works of Jones \cite{Jones} and Witten \cite{Witten1},
Reshetikhin and Turaev \cite{ReshTur}
developed the rigorous procedure based on the surgery representation of
3-manifolds, which enabled them to express the Chern-Simons partition
function through some link invariant. Then, using quantum 6-$j$ symbols,
Turaev and Viro constructed an invariant of triangulated manifolds equal
to the Chern-Simons partition function modulo squared
\cite{TV}. As was shown in ref. \cite{Piun}, it also coincides with the
Kauffman's $q$-spin network invariant \cite{Kauf}. The most general
framework for the Turaev-Viro construction was given in ref. \cite{Durh}.

Apart from this activity, the point of view exists that quantum
groups understood as automorphisms of quantum spaces \cite{Manin}
can provide us with a non-conradictory picture of the real world
including quantum gravity \cite{Majid}.
Now, what may be called $q$-deformed physics\footnote{We should adopt
the term $q$-deformed here as quantum
physics has already existed. Also, we say about $q$-deformed gauge theory
because the word ``quantum'' would be misleading in this context.}
is in a state of development.

This article is devoted to the $q$-deformation of lattice gauge
theory so that, in the limit $q\to 1$, the usual
``classical'' gauge symmetry is restored. We are using the approach dual
to the one based on quantized universal envelopping
algebras. We consider gauge variables taking values in a quantum gauge
group, {\em i.e.,} they are matrices with non-commuting elements.
The function algebra on a quantum group is a well defined object
\cite{SoiVak} and, for compact groups, there exists
the analog of the Peter-Weyl theorem
\cite{Woron}. Actually, we have all mathematical
tools at our disposal.

It appears that $q$-deformed gauge theory gives a
natural framework for the investigation of the Turaev-Viro invariant,
which is simply connected with the weak coupling limit of
a lattice partition function, when the deformation parameter is a root
of unity: $q=e^{i\frac{2\pi}{n}}$. Therefore, our model can be regarded
as a lattice regularization of 3-dimensional Chern-Simons theory. More
precisely, it contains both holomorphic and anti-holomorphic
Chern-Simons sectors. The obvious parallel with the relation between
conformal field theories  and 2-dimensional lattice models becomes more
clear if we consider a manifold with a connected boundary. Then, having
fixed all variables on the boundary and integrated inside, we obtain a
lattice regularization of the full WZNW model, both holomorphic and
anti-holomorphic sectors included. Appearing topological lattice gauge
theory is complementary to ones considered in ref. \cite{DijkW}.

Another exciting  application came from the Witten's result on the
equivalence between Chern-Simons theory and 3-$d$ quantum gravity
\cite{Witten}. Later
a number of papers appeared advocating a connection of the Turaev-Viro
partition function and its ``classical'' Ponzano-Regge \cite{PR}
counterpart with 3-$d$ euclidean gravity \cite{Ooguri}.
However, it is not all that should be done.
We have also to sum over base manifolds including different topologies.
This can, in principle, be performed within the simplicial approach to
gravity \cite{simgrav,B}. A using of triangulations assumes a piece-wise
linear appriximation. Fortunately, in 3-dimensions (as well as in two
and in contrast to higher dimensional cases) the notions of piece-wise
linear and smooth manifolds essentially coincide \cite{Nash}. We have a
strong reason to be sure that, in  the simplicial approximation, the
entropy of 3-manifolds can be correctly estimated. Although there is no
analytical progress in this direction, numerical simulations
\cite{simgrav} gave a lot of interesting results. It was shown, for
example, that the number of complexes of the spherical topology is
exponentially bounded as a function of a volume, ({\em viz.,} the number
of tetrahedra).

This paper has been written at the physical level of rigor.
We always try to establish  a connection with knot theory
and use pictures instead of formulas.
It makes most explanations more transparent without loss of
accuracy.

The paper is organized as follows. In Section~2 we define the $q$-deformed
lattice gauge model and develop an appropriate graphical technique to
work with it. Almost all ``classical'' formulas hold true, but one have
to be careful about an order of variables, framings of loops, etc.
In this section, we also establish correspondence with knot theory by
representing our partition function through the Jones polynomial of a
certain link. In Section~3 we define the Turaev-Viro invariant for cell
complexes and show how it can be interpreted.
General consideration is illustrated by a simple
example of lenses. In Section~4 we define a model which partition
function gives a generating function of all possible simplicial
complexes weighted with the Turaev-Viro invariant. In Appendix,
the \mbox{2-dimensional} version of our model is considered.

\newpage
\vspace*{4pc}

\stepcounter{chapter}

{\Large\bf q-deformed lattice gauge theory.}

\vspace{2pc}

Since the Wilson's famous paper \cite{Wilson}
lattice gauge theory has continuously
been drawing attention. This section is devoted to the generalization of
the Wilson's construction consisting in the $q$-deformation of a gauge
group $G$. The mathematical framework for this procedure is provided
by the Woronowicz approach
to quantum groups \cite{Woron}.

As usual, given a $d$-dimensional lattice,
a gauge variable $u_b\in G$ is attached to each its 1-dimensional bond $b$,
and the weight

\be
w_\beta(x_f)=\sum_R d_\R \chi_\R(x_f) e^{-\beta C_R}
\label{weight}
\ee
\ni
to each 2-dimensional face $f$. By a lattice we mean a cell
decomposition of a \mbox{$d$-dimensional} manifold (a polyhedron)
such that any
cell can enter in a boundary of another one only once, and every two
cells can border upon each other along only one less dimensional
cell. Simplicial complexes and their duals
obey this restriction by definition. In eq. (\ref{weight}),
$\sum_R$ is the sum over all irreps of the gauge group $G$;
$\chi_\R(x_f)$ is a character of an irrep $R$ ($x_f\in G$);
$d_R=\chi_\R(I)$ is the quantum dimension; $C_R$ is a
second Casimir and $\beta$ is a parameter. In this paper we consider
only compact groups and unitary finite dimensional representations.
Therefore, $R$ is always a discret index. We have chosen $w_\beta(x_f)$ in
such a form that it becomes the group $\delta$-function when $\beta\to
0$:

\be
w_0(x_f)=\delta(x_f,I)
\label{w0}
\ee

Arguments of the weights (\ref{weight}) are the ordered products of the
gauge variables along 1-$d$ boundaries $\partial f$ of corresponding
faces $f$:

\be
x_f=\prod_{k\in \partial f} u_k
\label{arg}
\ee

The partition function is defined by the integral

\be
Z_\beta=\int_G\prod_b du_b\: \prod_f w_\beta(\prod_{k\in\partial f}u_k)
\label{Z}
\ee
\ni
where $du_b$ is the Haar measure for the group $G$.

In the definition (\ref{Z}), two multiplications were used. The first
one is the usual matrix product in eq.(\ref{arg}). The second is the
tensor product. For classical groups the latter is commutative
and the order of
factors under the integral in eq. (\ref{Z}) is not important. The
quantization ($q$-deformation) makes the tensor product non-commutative
and gauge variables corresponding to the same bond have to be, in
principle, somehow ordered. It can be naturally interpreted as an
ordering of faces incident to this bond.
Depending on the dimensionality of a lattice three cases occure:\\
\ni
1) $d=2$. Every bond is a common boundary of exactly two faces, which
relative order should be unimportant.\\
\ni
2) If $d=3$, the natural cyclic order of faces incident to a bond is
always defined. By the Poincar\'{e} duality, every bond corresponds to a
dual face and {\it vice versa}. The above order is just the one of dual
bonds  around the boundary of a corresponding dual face.\\
\ni
3) If $d\geq 4$, there is no natural order of faces around a bond, which
forces the tensor product to be commutative.

Now, let us precise the above qualitative consideration. A character
$\chi_\R(x_f)$ is the trace of a matrix element, $D^R(x)$, of an irrep
$R$. Hence,

\be
\chi_\R(\prod_{j\in \partial f}u_j)=tr_\R \prod_{j\in\partial f}D^R(u_j)
\label{char}
\ee
\ni
The product on the r.h.s of eq. (\ref{char}) is taken with respect to
indices in a space, $V_R$, of an irrep $R$. As all gauge variables in
eq. (\ref{char}) are
different, the relative order of matrix elements in this case is not important.
Substituting eq. (\ref{char}) in eq. (\ref{weight}) and then in
(\ref{Z}) we get the partition function in the form

\be
Z_\beta=\int_G\prod_b du_b\: \sum_{\{R\}} \prod_f
d_{R_f}\: e^{-\beta C_{R_f}}\; tr_{\R_f}\bigg[
\prod_{k\in\partial f}D^{R_f}(\stackrel{\ell_{kf}}{u_k})\bigg]
\label{Z2}
\ee
\ni
where integers $\ell_{kf}$ above variables show their relative order
with respect to the tensor product.
In what follows, matrix elements, $D^R_{ab}(u),\ a,b\in V_R$, will be
basic objects. We need to calculate integrals of their products. Using
the Clebsch-Gordan coefficients, $\langle aRbS\vert cT\rangle$, (which
are elements of a unitary matrix performing the decomposition of the
tensor product $V_R\otimes V_S$ in irreps) we can decompose the product
of two matrix elements as follows

\be
D^R_{aa'}(x)D^S_{bb'}(x)=\sum_{T;\,c,c'\in V_T}\langle aRbS\vert cT\rangle
\overline{\langle a'Rb'S\vert c'T\rangle}D^T_{cc'}(x)
\label{decomp}
\ee
\ni
Applying eq. (\ref{decomp}) successively we can, in principle, reduce
an arbitrary group integral to the trivial one

\be
\int_G du\: D^R(x)=\delta^{R,0}
\label{trivint}
\ee
\ni
({\it i.e.,} the integral (\ref{trivint}) equals 1 for the trivial
representation, $D^0(x)\equiv 1$, and 0, otherwise). Eqs. (\ref{decomp}) and
(\ref{trivint}) can be regarded as a formal definition of the integral over a
quantum group. An answer is always a combination of Clebsch-Gordan
coefficients.

In order to make our formulas more transparent, let us adopt the
graphical representation based on the Reshetikhin's 3-valent colored
graphs \cite{Resh,Kiresh}.
In what follows, small letters will stay for indices in spaces
of representations, the bar being the conjugation. Capitals will
numerate irreps.

We use the following diagammatic elements:
\par
\medskip
\ni
The $\delta$-function in a representation space, $a,a'\in V_R$,
\be
\delta_{aa'}=\begin{array}{c} a'\\ | \\ \uparrow\\a\end{array}
\ee
\ni
The Clebsch-Gordan coefficients,

\be
\langle aR\,bS\vert cT\rangle=
\begin{array}{rcl} &cT&\\ &|& \\&& \\ aR&&bS\end{array}
\hspace{1cm}
\overline{\langle aR\,bS\vert cT\rangle}=
\begin{array}{rcl} aR&&bS\\ & & \\ &|&\\&cT& \end{array}
\ee
\ni
The conjugation in a representation space,

\be
\langle aR\,\overline{a}R\vert 0\rangle\sqrt{d_R}=\begin{array}{ccc}
&&\\&&\\aR& &aR\end{array}
\hspace{1cm}
\langle \overline{a}R\,aR\vert 0\rangle\sqrt{d_R}=\begin{array}{ccc}
aR&&aR\\&&\\&&\end{array}
\ee
\ni
The matrix element
\be
D^R_{ab}(u)=\begin{array}{c} bR\\ \uparrow \\ \fbox{$u$}\\ \uparrow
 \\aR\end{array}
\ee
\ni
having the following properties

\[\rule[-1.5cm]{0cm}{3cm}
\begin{array}{ccc}
bR&&aR\\
\uparrow&&\downarrow\\
\fbox{$x^{+}$}&=&\fbox{$x$}\\
\uparrow&&\downarrow\\
aR&&bR
\end{array}\hspace{2cm}
\begin{array}{ccc}
&& | \\
|&&\fbox{$x$}\\
\fbox{$xy$} &=& | \\
|&&\fbox{$y$}\\
&&|
\end{array}
\]

\be\rule[-1.5cm]{0cm}{3cm}
\fbox{$u$}\ \ \fbox{$u$}\ = \hspace{3cm} \fbox{$I$}\ =\ \uparrow
\label{proper}
\ee
\ni
The character

\be
\chi_\R(u)=\sum_{a\in V_R}\ \fbox{$u$}
\ee
\ni
The quantum dimension of $V_R$

\be
d_R=\chi_\R(I)=\sum_{a\in V_R}\hspace{1cm}
\ee
\ni

\medskip

Matrix elements are ordered from the left to the right

\be
\int du\: D^{R_1}_{a_1a'_1}(u)\ldots D^{R_n}_{a_na'_n}(u)=
\int du
\begin{array}{cccc}
\scr{a'_1R_1}&\scr{a'_2R_2}&&\scr{a'_nR_n}\\
|&|&&|\\
\fbox{$u$}&\fbox{$u$}&\cdots\cdots&\fbox{$u$}\\
|&|&&|\\
\scr{a_1R_1}&\scr{a_2R_2}&&\scr{a_nR_n}
\end{array}
\label{bigint}
\ee

Now, let us give several simple examples. Eq. (\ref{decomp}) looks like

\be\rule[-2cm]{0cm}{4cm}
\fbox{$u$}\fbox{$u$}\ =\sum_{T;\,c,c'\in V_T}\ \fbox{$u$}
\ee

\ni
The orthogonality of matrix elements takes the following graphical form

\be\rule[-1.5cm]{0cm}{3cm}
\int du\; D^R_{aa'}(u)D^S_{bb'}(u)=\frac{\delta^{\scs{R,S}}}{d_R}
\ee

\ni
The $\delta$-function acts as

\be\rule[-1.5cm]{0cm}{3cm}
\int dx\; \delta(u,x)f(x)=\int dx\sum_R d_R\;\fbox{$u$}\ \ \fbox{$x$}\;
\sum_S \fbox{$x$}\ \ \fbox{$f^{\scs{S}}_{ab}$}=\sum_S\fbox{$u$}\ \
\fbox{$f^{\scs{S}}_{ab}$}=f(u)
\ee

\ni
{}From the equality

\be\rule[-2cm]{0cm}{4cm}
\fbox{$u$}\fbox{$u$}\ =\sum_{T;\,c,c'\in V_T}\ \fbox{$u$} \
=\sum_{T;\,c,c'\in V_T}\ \fbox{$u$} \ = \
\fbox{$u$} \fbox{$u$}
\label{permut}
\ee

\ni
it follows that we can permute matrix elements freely as far as
corresponding links remain equivalent. Eq. (\ref{permut}) allows us to
give another more geometrical presentation of our model in the $d=3$
case. Characters, which are traces
of products of matrix elements, can be imagined as loops going around
dual bonds. They intersect dual faces at points (associated with matrix
elements), which can walk freely on
these faces. For lattices an appearing link is always trivial and a
relative order of matrix elements is unimportant! However,
it will not be the case for general cell complexes considered in the next
section. So, we have even too much symmetry, as {\it a priori}
we needed only the cyclic one.
If $d\geq4$, we return essentially to ordinary gauge theory,
since there such loops are forced to be mutually penetrable. However, we
can get non-trivial generalizations also in this case using triangular
Hopf algebras.
In what follows, we shall limit ourselves to the most interesting $d=3$
case.

\medskip
If in eq. (\ref{weight}) $\beta > 0$, the partition function
(\ref{Z}) is well defined for any finite cell decomposition of a
closed oriented\footnote{ For non-oriented complexes we would have to use
orthogonal groups} 3-manifold. However, the weak coupling limit,
$\beta=0$, is non-singular, only if the deformation parameter is
a root of unity, $q^n=1$.
In this case, gauge field is just a pure gauge. We have no
dynamical degrees of freedom and the partition function can depend only
on a topology of a lattice.

For classical gauge groups the self-consistency of the model follows
from the
Peter-Weyl theorem stating that the algebra of regular functions on a
compact group is isomorphic to the algebra of matrix elements of finite
dimensional representations. The quantum version of this theorem was proved
in refs. \cite{Woron} for $\vert q \vert < 1$. In this case there is
the one-to-one correspondence between representations of $SU_q(N)$ and
$SU(N)$, and the notion of the matrix elements is naturally generalized.
The condition $q=e^{2\pi i/n}$ changes the situation and the analysis of
refs. \cite{Woron} is not valid anymore.
For concreteness, let us consider the simplest
$SU(2)$ case. The theory of representations of the
quantized universal enveloping algebra\footnote{ Which is dual to
$SU_q(2)$ \cite{Rosso}.}
${\cal U}_q(SL(2))$, when
$q^n=1$, was given in refs. \cite{repsu2,Keller}. For the sake of
comleteness, let us repeat main facts, which we shall use, as they were
formulated in ref. \cite{Keller}.

All highest weight irreps $\rho_j$ of
$\ {\cal U}_q(SL(2))$, when $q^n=1$, fall into two classes:\\
a) dimension of $\rho_j,\ dim(\rho_j)\ <M$, where
$M=\left\{\begin{array}{ll}
n/2&\mbox{$n$ even}\\
n&\mbox{$n$ odd}
\end{array}\right.$\\
These irreps are numbered by two integers $d$ and $z$, $\langle
d,z\rangle$, where $d=dim(\rho_j)$, and the highest weights are

\be
j=\frac{1}{2}(d-1)+\frac{n}{4}z
\label{highw}
\ee
\ni
b) $dim(\rho_j)=M$. In this case irreps $I^1_z$ are labeled by a complex
number $z\in {\rm C}\backslash\big\{{\rm Z}+\frac{2}{n}r\mid 1\leq r\leq
M-1\big\}$ and have the highest weights

\be
j=\frac{1}{2}(M-1)+\frac{n}{4}z
\ee

There are also indecomposable representations which are not
irreducible but nevertheless cannot be expanded in a direct sum of
invariant subspaces. They are labeled by an integer $2\leq p\leq M$ and
the complex number $z$: $I^p_z$. Their dimension $dim(I^p_z)=2M$.

In ref. \cite{Keller} the following important for us facts were
established:\\
1) If $n\geq 4$, irreps $\langle d,0\rangle$ are unitary only for even
$n$.\\
2) Representations of the type $I^p_z,\ 1\leq p\leq M$ form a two sided
ideal in the ring of representations ({\em i.e.}, if at least one of
them appears in a tensor product, then all representations in the
decomposition will be of this type). Their quantum dimension vanishes:
\(dim_q(I^p_z)=\left\{\begin{array}{cc} {[M]}&,p=1\\ {[2M]}&,p\geq
2\end{array}\right\}=0\), where $[x]=\frac{q^x-q^{-x}}{q-q^{-1}}$.\\
3) For the tensor product of two irreps the following formula takes
place:

\[
\langle i,z\rangle\otimes\langle j,w\rangle=
\bigg(\bigoplus_{k=\vert i-j\mid+1;+3;+5,\ldots}
^{\min(i+j-1,2M -i-j-1)}\langle k,z+w\rangle\bigg)\oplus
\]\be
\bigg(\bigoplus_{\ell=r,r+2,r+4,\ldots}^{i+j-M}
I^{\ell}_{z+w}\bigg)
\label{tenprod}
\ee
\ni
where $r=\left\{\begin{array}{ll}
1&\mbox{$i+j-M$ odd}\\
2&\mbox{otherwise}
\end{array}\right.$\\

Eq. (\ref{tenprod}) means that the class of representations $\langle
d,z\rangle$
and $I^p_z$ with $z=0$ forms a ring with respect to the tensor product
and all other representations form an ideal. Because of eq.
(\ref{trivint}), all sums over irreps in the definition of the
partition function (\ref{Z2}) are actually truncated to irreps
from the ring. On the other hand, $I^p_0$ representations does not
contribute to $Z_\beta$, because their quantum dimensions vanish. So,
without loss of self-consistency, we can use in eq. (\ref{weight}) the
space of functions spanned only by irreps $\langle d,0\rangle,\ 1\leq
d\leq n/2-1$, for even $n\geq 4$. Their highest weights are in the
one-to-one correspondence with ones for $| q | < 1$. Their matrix
elements have non-singular limit as $q\to e^{2\pi i/n}$ and the model
reminds in some respects lattice gauge theory with a finite gauge group.
Of course, the limit $q\to e^{2\pi i/n}$ has to be taken before $\beta
\to 0$. From now on, we shall consider only this case.

In order to establish a connection with the Turaev-Viro invariant, let us
consider lattices dual to simplicial compexes. Their 1-skeletons are
4-valent graphs and exactly 3 faces are incident to each bond.
Hence, we have the integral of 3 matrix elements attached to every triangle
in a simplicial complex:

\be\rule[-2cm]{0cm}{4cm}
\int du\; D^{R_1}_{a_1b_1}(u) D^{R_2}_{a_2b_2}(u)
D^{R_3}_{a_3b_3}(u) \equiv \frac{1}{d_{R_3}}\hspace{4cm}
\label{3j}
\ee
\ni
The right hand side of eq. (\ref{3j}) is the product of two 3-$j$
symbols. Summing over lower indices we get a Racah-Wigner 6-$j$
symbol

\be\rule[-2cm]{0cm}{4cm}
\left\{\ba{ccc}R_1&R_2&R_3\\R_4&R_5&R_6\ea\right\} \equiv
\frac{1}{d_{R_5}\sqrt{d_{R_3}d_{R_6}}}\hspace{4cm}
\label{6j}
\ee

\ni
inside each
tetrahedron of a simplicial complex. Representation indices, $R_f$, are
attached to its 1-simplexes, $f$, ({\em i.e.} faces of an original lattice).
The partition function $Z_{\beta\to 0}$ can be written then in the Turaev-Viro
form

\be
Z_0= \sum_{\{R\}} \prod_f d_{R_f}\:  \prod_t
\left\{\ba{ccc}R_{t_1}&R_{t_2}&R_{t_3}\\R_{t_4}&R_{t_5}&R_{t_6}\ea\right\}
\label{TuV}
\ee
\ni
where the indices $t_1,\ldots,t_6$ stay for six edges of a $t$'th
tetrahedron.

The interpretation of the Turaev-Viro invariant as the weak coupling
limit of lattice $q$-gauge theory makes many proofs more
transparent. Actually, all standard technique is applicable through eqs.
(\ref{trivint}) - (\ref{permut}) and the invariance of the Haar measure.
However, there is another useful formulation of the model which
enables us to establish a connection with knot theory directly.

An expert should notice that eq. (\ref{permut}) is of one of the forms of
the so-called $Rtt=ttR$ equation in the theory of integrable models.
Then, we can try to substitute the $R$-matrix

\be\rule[-1cm]{0cm}{2cm}
R\ = \hspace{4cm}
\ee

\ni
for matrix elements in eqs. (\ref{trivint}) - (\ref{permut}). Eq.
(\ref{permut}) becomes the Yang-Baxter equation written in the form

\be
\rule[-1.5cm]{0cm}{3cm}
\mbox{\Large =}
\ee
\ni
The intergral (\ref{bigint}) is substituted by the following tangle (in
the terminology of ref. \cite{ReshTur})

\be
\frac{1}{| G |}\sum_S d_S\
\begin{array}{cccc}
\scr{a'_1R_1}&\scr{a'_2R_2}&&\scr{a'_nR_n}\\
|&|&&|\\
&&\cdots&\\
|&|&&|\\
\scr{a_1R_1}&\scr{a_2R_2}&&\scr{a_nR_n}
\end{array}
\label{tangle}
\ee
where $\sum_S$ is the sum over all irreps $S$ attached to the loop
pinching the bunch of vertical lines; $d_S$ are their quantum
dimensions and

\be
| G | = \delta(I,I)=\sum_S (d_S)^2
\label{qrank}
\ee
\ni
is the quantum rank of a gauge group. In the $SU_q(2),\
q=e^{i\frac{2\pi}{n}}$ case, we find

\be
|SU_q(2)|=\frac{n}{2\sin^2(\frac{\pi}{n})}
\ee

The analog of eq. (\ref{decomp})
takes the form

\be\rule[-1.5cm]{0cm}{3cm}
=\ \sum_T\hspace{2cm}=\ \sum_T\hspace{1cm}
\label{pinch}
\ee
\ni
where the unitarity of 3-$j$ symbols was used. Using eq. (\ref{pinch})
subsequently we can reduce an arbitrary tangle (\ref{tangle})
to the simplest one (the analog of the basic integral (\ref{trivint}))

\be\rule[-1.5cm]{0cm}{3cm}
\frac{1}{|G|}\sum_S d_S\hspace{1.5cm}=\ \delta_{R,0}
\label{trivbundle}
\ee

Eq. (\ref{trivbundle}) can be checked directly or extracted from formulas
of ref.
\cite{ReshTur}. In what follows, we shall need also the analog of the
Haar measure invariance

\be
\int du\;f(ux,x)\,=\int du\;f(u,x)
\ee
\ni
which can be represented graphically as follows

\[\rule[-1.5cm]{0cm}{4cm}
\frac{1}{|G|}\sum_S d_S\hspace{4cm}=\ \frac{1}{|G|}\sum_S d_S\hspace{4cm}=
\]
\be\rule[-1.5cm]{0cm}{3cm}
=\  \frac{1}{|G|}\sum_S d_S\hspace{8cm}
\ee

We are now in position to formulate the partition function $Z_0$ as a
link invariant. To every cell decomposition of a 3-manifold, we put into
correspondence a link $L$ constructed of $\alpha_1+\alpha_2$ unknotted
loops with zero framings ($\alpha_i$ is the number of $i$-dimensional
cells in the cell complex). $\alpha_2$ loops go along boundaries of
2-cells; $\alpha_1$ loops go around 1-cells pinching loops attached to
incident faces. Every loop carries a representation of a gauge group.
Having calculated the Jones polynomial
$J^{T_1\ldots T_{\alpha_2}}_{S_1\ldots S_{\alpha_1}}(L)$
for the link $L$,
we get the partition function $Z_0$ summing over all representations:

\be
Z_0=\sum_{\{T\}}\sum_{\{S\}}  \prod_{f=1}^{\alpha_2}d_{T_f}
\prod_{b=1}^{\alpha_1}\frac{d_{S_b}}{|G|}\:
J^{T_1\ldots T_{\alpha_2}}_{S_1\ldots S_{\alpha_1}}(L)
\label{Z3}
\ee

The reader should have noticed the analogy with the Reshetikhin-Turaev
construction of the Witten's Chern-Simons invariant \cite{ReshTur}.
However, in our case, the link $L$ is not connected directly with a Dehn
surgery of a manifold.

To conclude this section, let us note that the
partition function $Z_0$ in the form (\ref{Z3}) is manifestly self-dual
with respect to the Poincar\'{e} duality of complexes.

\vspace*{4pc}

\stepcounter{chapter}
{\Large\bf Invariants of cell complexes}

\vspace*{2pc}

In the previous section we have defined the weak coupling partition
function  $Z_0$ of 3-$d$ $q$-gauge theory as the integral

\be
Z_0(C)=\int_G \prod_{i=1}^{\alpha_1} du_i\:
\prod_{j=1}^{\alpha_2}\delta(\prod_{\sigma^1_k\in\partial\sigma^2_j}
\stackrel{\ell_{jk}}{u_k},I)
\label{Z0}
\ee
\ni
where $C$ is a cell decomposition of a closed oriented 3-manifold $M^3$
consisting of $\alpha_k$ $k$-dimensional cells:
\be
C=\bigcup_{n=1}^{\alpha_0} \sigma^0_n
\bigcup_{i=1}^{\alpha_1} \sigma^1_i
\bigcup_{j=1}^{\alpha_2}\sigma^2_j
\bigcup_{m=1}^{\alpha_3}\sigma^3_m
\ee
\ni
Let us remind that for 3-manifolds the Euler character
$\chi=\alpha_0-\alpha_1+\alpha_2-\alpha_3=0$.
To each 1-cell $\sigma^1_i$, we put into correspondence a gauge
variable $u_i\in G$, and to each 2-cell $\sigma^2_j$, a
$\delta$-function. The argument of the $\delta$-function repeats a
form of the boundary of the 2-cell, $\partial\sigma^2_j$, written down
multiplicatively. Integers $\ell_{jk}$ show the relative order
of variables attached to the same 1-cell, $\sigma^1_k$. This order is
defined by 2-cells in the co-boundary, $\delta\sigma^1_k$, of
the 1-cell (which can be interpreted as the boundary of a dual 2-cell,
$\tilde\sigma^2_k$, in the dual complex, $\tilde C$).

It is known that any 3-dimensional
oriented manifold can be represented by a cell complex having only one
0-cell, $\sigma^0$, one 3-cell, $\sigma^3$
and the equal number, $\nu$, of 1-cells, $\sigma^1_i,\ i=1,\ldots,\nu$,
and 2-cells, $\sigma^2_j,\ j=1,\ldots,\nu$. An arbitrary cell
decomposition of the manifold can be
transformed into this canonocal form by a sequence of topology preserving
deformations (moves). The simplest sufficient set of such moves consists
of:\\
\ni
(1) The shrinking of a 1-cell, $\sigma^1_i$, with the
identification of 0-cells at its ends:

\be
\sigma^1_i\bigcup_{\sigma^0_k\in\partial\sigma^1_i}
\sigma^0_k\to \sigma^0_*
\ee
\ni
This move corresponds to the shift of variables
under the integral (\ref{Z0}) eliminating a variable $v$ attached to
$\sigma^1_i$:

\be
\int \prod_{i=1}^n du_i \int dv\: f(u_1v,\ldots,u_nv)=
\int \prod_{i=1}^n du_i\: f(u_1,\ldots,u_n)
\label{varout}
\ee
\ni
where $f$ is an arbitrary, in general, function having no additional
dependence on $u_1,\ldots,u_n$. In the quantum case, the shift
(\ref{varout}) is possible always, when it takes place under the classical
integral. Indeed, it is just a combination of eqs. (\ref{proper}) and
(\ref{permut}) not changing a link.
See an example in Fig.~1.\\

\ni
(2) The shrinking of a 2-cell, $\sigma^2_j$, disjoining
two 3-cells which have bordered upon each other along the 2-cell.

\be
\sigma^2_j\bigcup_{\sigma^1_k\in\partial\sigma^2_j}
\sigma^1_k \bigcup_{\sigma^0_i\in\partial\sigma^1_k}\sigma^0_i
\to \sigma^0_*
\ee
\ni
For the integral (\ref{Z0}), it implies the using of the following
$\delta$-function property:

\be
\int \prod_{i=1}^n du_i\: f(u_1,u_2,\ldots,u_n)\:
\delta(u_1u_2\ldots u_n,I)=
\int \prod_{i=1}^n dv_i\: f(v_n^{-1}v_1,v_1^{-1}v_2,\ldots,v_{n-1}^{-1}v_n)
\label{deltout}
\ee
\ni
and a subsequent absorption of the $v$-variables as in eq. (\ref{varout}).
All variables in the argument of the $\delta$-function are assumed to be
different. Eq. (\ref{deltout}) has the following simple interpretation
(see Fig.~2). The $u$-variables correspond to 1-cells forming the
boundary of the 2-cell, the $\delta$-function is associated with. They
define a sequence of 0-cells at their ends. Let us put a new 0-cell
inside the 2-cell and connect it with other 0-cells by 1-cells
$\sigma^1_{v_1},\ \sigma^1_{v_2},\ldots,\sigma^1_{v_n}$. Then we can
easily express $u$ cells through $v$ ones:
$\sigma^1_{u_k}=-\sigma^1_{v_{k-1}}+\sigma^1_{v_k}$.\\

\ni
(3) Using moves (1) and (2) we can get a 2-cell, $\sigma^2_j$, without a
1-dimensional boundary. It is exactly a case of the sum of complexes:
$C=C_1\#C_2$, if the 2-cell divides two 3-balls glued by their spherical
boundaries, $S^2\cong\sigma^0\cup\sigma^2_j$. If the 2-cell $\sigma^2_j$
disjoins two 3-cells $\sigma^3_1\in C_1$ and $\sigma^3_2\in C_2$, we can
always unify them

\be
\sigma^3_1\cup\sigma^2_j\cup\sigma^3_2\cong\sigma^3_*
\ee
\ni
Hence, we get

\be
Z_0(C)=\frac{Z_0(C_1)Z_0(C_2)}{|G|}
\ee
\ni
and the quantity

\be
{\cal I}(C)={Z_0(C)}/{|G|^{\alpha_3-1}}
\label{invar}
\ee
\ni
is a topological invariant. With the normalization (\ref{invar})

\be
{\cal I}(S^3)=1
\ee
and

\be
{\cal I}(C_1\#C_2)={\cal I}(C_1){\cal I}(C_2)
\ee

For lattices considered in the previous section, all links were
unframed. However, on applying the above moves we can obtain non-trivial
framings and, in real calculations, we have to follow them carefully.
Practically, it is convenient to use the ribbon graph representation by
Reshetikhin and Turaev \cite{ReshTur}. But the manipulations
with gauge variables in eqs. (\ref{varout}), (\ref{deltout}) are not
sensitive to framings, as they use only formal properties of matrix
elements and the group measure.
\medskip

If a complex has been
transformed into its canonical form:

\be
C=\sigma^0
\bigcup_{i=1}^{\nu} \sigma^1_i
\bigcup_{j=1}^{\nu}\sigma^2_j\cup
\sigma^3
\ee
\ni
we can put into correspondence with each its 1-cell, $\sigma^1_i$, a
generator of the fundamental group $g_i\in\pi_1(C)$. Each 2-cell gives a
defining relation for $\pi_1(C)$:

\be
\Gamma_j=\prod_{\sigma^1_k\in\partial\sigma^2_j}g_k=I
\label{defrel}
\ee

If a gauge group $G$ is a classical finite group, the partition function
(\ref{Z0}) is well defined (after the substitution of the sum
$\sum_{u\in G}$ for $\int du$). Then the invariant ${\cal I}(C)$  can be
interpreted as the rank of the homomorphism of the fundamental group
into the gauge one: $\pi_1(C)\to G$ \cite{B}. In other words, ${\cal
I}(C)$ is equal to the number of representations of $\pi_1(C)$ by
elements of $G$. Hence, it is an integer.

Let us try to find a similar interpretation in the quantum case:
$G=SU_q(N),$ $q^n=1$. To do it, let us consider an action of $\pi_1(C)$
on  a universal covering $\hat{C}$ ({\em i.e.}
$\pi_1(\hat{C})=1$). Let $\pi_1(C)$ act freely permuting cells of $\hat{C}$.
All 0-cells, $\hat{\sigma}^0_i\in\hat{C}\ (i=1,\ldots,|\pi_1(C)|)$, form
an orbit with respect to this action and can be formally obtained acting
by elements of the fundamental group, $h_i\in\pi_1(C)$, on
an arbitrarily chosen cell $\hat{\sigma}^0_1$:

\be
\hat{\sigma}^0_i=h_i\hat{\sigma}^0_1
\label{basis0}
\ee
\ni
It makes the set of 0-cells a $\pi_1(C)$-module.
If $g_kh_i=h_j\in\pi_1(C)$, we find

\be
g_k\hat{\sigma}^0_i=\hat{\sigma}^0_j
\ee

All 1-cells, $\hat{\sigma}^1_{k,i}\in\hat{C}$
$(k=1,\ldots,\nu;$ $i=1,\ldots, |\pi_1(C)|)$,
can be identified with formal differences (or pairs) of 0-ones using the
boundary operator

\be
\partial\hat{\sigma}^1_{k,i}=(g_k-1)\hat{\sigma}^0_{i}
=(h_{j}-h_{i})\hat{\sigma}^0_1
\label{1cells}
\ee
\ni
Eq. (\ref{1cells}) induces an action of $\pi_1(C)$ on 1-cells of
$\hat{C}$.

Owing to the Poincar\'{e} duality, we can put 3-cells,
$\hat{\sigma}^3_i\in\hat{C}\ (i=1,\ldots,|\pi_1(C)|)$, into
correspondence with 0-ones $\hat{\sigma}^0_i$ and numerate them formally
with the same $\pi_1(C)$ elements. 2-cells,
$\hat{\sigma}^2_{k,i}\in\hat{C}$
$(k=1,\ldots,\nu;$ $i=1,\ldots, |\pi_1(C)|)$,
can be coded using the co-boundary
operator:

\be
\delta\hat{\sigma}^2_{k,i}=\hat{\sigma}^3_{j}-\hat{\sigma}^3_{i}
\label{2cells}
\ee

However, $\pi_1(C)$ itselt can act on 3-cells of $\hat{C}$
non-trivially. Of course, after a projection onto $C$,
this action has to coincide with the one
following from the Poincar\'{e} duality. We
shall call ``untwisted'' manifolds for which these two actions of
$\pi_1(C)$ coincide already on $\hat{C}$. And, it is this ``twisting''
that is lost in the invariant constructed with classical groups.

Any element of $\pi_1(C)$ can be written as a combination of generators
{\em modulo} defining relations. If $h_i=g_{i_1}g_{i_2}\ldots
g_{i_{\mu_i}}$, we write formally

\be
\hat{\sigma}^3_i=g_{i_1}\otimes g_{i_2}\otimes\ldots
\otimes g_{i_{\mu_i}}\hat{\sigma}^3_1
\label{basis}
\ee
Let us denote $G_k$ a representation
of a generator $g_{k}\in \pi_1(C)$ acting on the basis (\ref{basis}):
$G_k\hat{\sigma}^3_i=\hat{\sigma}^3_j$. Then $G_k$ can be realized as a
tensor operator with respect to the matrix multiplication of generators

\[
G_k\hat{\sigma}^3_i
=\big(g_{k_1}\otimes g_{k_2}\otimes\ldots
\otimes g_{k_{\mu_k}}\big)\otimes
\big(g_{i_1}\otimes g_{i_2}\otimes\ldots
\otimes g_{i_{\mu_i}}\big)\hat{\sigma}^3_1=
\]\be
=g_{j_1}\otimes g_{j_2}\otimes\ldots
\otimes g_{j_{\mu_j}}\hat{\sigma}^3_1=\hat{\sigma}^3_j
\label{Gact}
\ee
Eqs. (\ref{basis}), (\ref{Gact}) can be given the following rather
natural interpretation. The basis (\ref{basis}) is dual to
(\ref{basis0}) and we used the ``dual'' multiplication to write it down.
We do not pay attention to representation indices of generators; they can be
partially summed over, partially open.
In eqs. (\ref{basis}), (\ref{Gact}), every generator itself
can be identified with a 2-cell of $\hat{C}$ in virtue of
eq. (\ref{2cells}). On the other hand, its action on representation indices
corresponds to a 1-cell due to eq. (\ref{1cells}).

As an operator $G_k$ can connect
non-incident cells, it should be, in general, a tensor product of a
number of generators. We demand that it acts as a tensor operator on
representation indices (omitted in eq. (\ref{basis})) so that
the space spanned by (\ref{basis}) becomes a $\pi_1(C)$-module.
It gives us a number of equalities, namely a representation of the
defining relations (\ref{defrel}).
We  do not demand the $\pi_1(C)$-module to bear any
additional algebraic structures. For example, we do not need it to be a
Hopf algebra or something else.

The invariant (\ref{invar}) can be interpreted now as a ``quantum
rank'' of the representation of the above defined $\pi_1(C)$-module
by elements
of $SU_q(N)$, $q^n=1$. Indeed, the only thing we need to match is the
defining relations, in other respects eq. (\ref{Gact}) is a tensor
representation of a free group.

\medskip

In order to gain some experience, let us consider in details the simplest
example of the lens spaces: $L^3_p(k)=S^3/Z_p$. If $S^3$ is realized in
the 2-d complex space $(z_1,z_2)$ by the equation $|z_1|^2+|z_2|^2=1$,
$L^3_p(k)$ is obtained  identifying points connected by the $Z_p$
shifts:
$(z_1,z_2)\to(z_1e^{i\frac{2\pi}{p}},z_2e^{i\frac{2\pi}{p}k})$, $p$ and
$k$ are mutually prime. $\pi_1(L^3_p(k))=Z_p$ with one generator $g$
obeying the defining relation $g^p=1$. The cell decomposition of
$L^3_p(k)$ has one cell in each dimension:
$L^3_p(k)\cong\sigma^0\cup\sigma^1\cup\sigma^2\cup\sigma^3$ with
the boundaries and co-boundaries:

\be
\ba{ll}
\partial\sigma^0=0&\delta\sigma^0=0\\
\partial\sigma^1=0&\delta\sigma^1=p\sigma^2\\
\partial\sigma^2=p\sigma^1&\delta\sigma^2=0\\
\partial\sigma^3=0&\delta\sigma^3=0
\ea
\ee

The universal covering $\widehat{L^3_p(k)}$ has $p$ cells in each
dimension:

\be
\widehat{L^3_p(k)}\cong
\bigcup_{i=0}^{p-1}\sigma^0_i
\bigcup_{i=0}^{p-1}\sigma^1_i
\bigcup_{i=0}^{p-1}\sigma^2_i
\bigcup_{i=0}^{p-1}\sigma^3_i
\ee
$g$ acts on them as

\be
g\hat{\sigma^0_i}=\hat{\sigma}^0_{i+1}\hspace{1cm}
g\hat{\sigma^1_i}=\hat{\sigma}^1_{i+1}\hspace{1cm}
g\hat{\sigma^2_i}=\hat{\sigma}^2_{i+k}\hspace{1cm}
g\hat{\sigma^3_i}=\hat{\sigma}^3_{i+k}
\ee
where the indices have to be taken {\bf mod} $p$.

The corresponding (co) boundaries are

\be
\ba{ll}
\partial\hat{\sigma}^0_i=0&
\delta\hat{\sigma}^0_i=(g-1)\hat{\sigma}^1_i\\
\partial\hat{\sigma}^1_i=(g-1)\hat{\sigma}^0_i&
\delta\hat{\sigma}^1_i=(1+g^k+g^{2k}+\ldots+g^{k(p-1)})\hat{\sigma}^2_i\\
\partial\hat{\sigma}^2_i=(1+g+g^2+\ldots+g^{p-1})\hat{\sigma}^1_i
&\delta\hat{\sigma}^2_i=(g^k-1)\hat{\sigma}^3_i\\
\partial\hat{\sigma}^3_i=(g^k-1)\hat{\sigma}^2_i
&\delta\hat{\sigma}^3_i=0
\ea
\ee

The basis (\ref{basis}) in this case is given by
\newcommand{\n}{\alpha}
\be
\hat{\sigma}^3_i=\underbrace{g_{\n_1\n_{k+1}}\otimes
g_{\n_2\n_{k+2}}\otimes\ldots \otimes
g_{\n_{i}\n_{k+i}}}_{i{\rm -times}}
\hat{\sigma}^3_0
\ee
The generator,
$G=\underbrace{g_{\n_1\n_{k+1}}\otimes\ldots\otimes
g_{\n_k\n_{2k}}}_{k{\rm -times}}$,
acts on it as follows

\be
G\hat{\sigma}^3_i=g_{\n_1\n_{k+1}}\otimes
g_{\n_2\n_{k+2}}\otimes\ldots g_{\n_{i+k}\n_{i+2k}}
\hat{\sigma}^3_0=\hat{\sigma}^3_{(i+k)\bmod p}
\ee
The defining relation has the form

\be
\underbrace{g_{\n_1\n_{k+1}}\otimes
g_{\n_2\n_{k+2}}\otimes\ldots \otimes g_{\n_{p-k+1}\n'_1} \otimes
\ldots \otimes g_{\n_{p}\n'_{k}}}_{p{\rm -times}}
=\delta_{\n_1,\n'_1}\ldots\delta_{\n_k,\n'_k}
\label{repdefrel}
\ee
where all numbers are taken {\bf mod} $p$ and all repeated indices are
assumed to be summed over; non-paired indices are open.

To get the invariant, we have to substitute matrix elements for the
generators on the l.h.s. of eq. (\ref{repdefrel}), take the $k$-fold
quantum trace and sum over all irreps of a gauge group.
However, we have to be carefull with framings. To put them correctly, we
can use, for example, the interpretation given in the previous section
to index loops. As they bound 2-cells and our complexes are oriented, we
can use ribbon loops with black and white sides, a black one always turned
toward the inside of a 2-cell, the white one, outward. Actually, it is
the most natural and simple way to fix all framings. From this
consideration, one can see that framings, in general, are connected with
torsion elements in a complex.

The untwisted lenses $L^3_p(1)$ are the most trivial example. Here,
$G=g_{\alpha_1\alpha_2}$, and the defining relation (\ref{defrel}) is
just $g^p=I$. The simplest way to calculate ${\cal I}(L^3_p(1))$ is to
reduce it to the Witten's Chern-Simons invariant
$I_{CS}(L^3_p(1))$ via a surgery representation \cite{ReshTur}.
Let us use the representation (\ref{Z3}). The
corresponding link is shown in Fig.~3. We have one unframed
loop (which would be a group integral in the gauge model).
The second loop is twisted $p$ times around the first one
and $p$ framed. Using the Kirby $\beta$-move \cite{Kirby},
we can disjoin them. This move changes links not changing a manifold
obtained by corresponding surgeries. Let us remind shortly how it is
performed. Given two loops $\alpha$ and $\beta$ with framings $a$ and
$b$, respectively,  we change $\alpha$ for another loop $\alpha'$
constructed as follows. Let $\beta_1$ be a parallel of $\beta$
twisted $b$ times around it (as two edges of a ribbon representing
$\beta$). Then $\alpha'=\alpha\cup\beta_1$ consists of a loop $\alpha$
cut at some point and connected by parallel wires to $\beta_1$.
The framing of $\alpha'$ is equal to $a+b+\ell(\alpha,\beta)$, where
$\ell(\alpha,\beta)$ is a linking coefficient of $\alpha$ and $\beta$
agreed with the framings. $\beta$ is not changed by the move.

For $L^3_p(1)$ we can easily get two disjoint loops with framings $p$
and $-p$ as shown in Fig.~3 and find

\be
{\cal I}(L_p^3(1))= |\hbox{\em I}_{CS}(L_p^3(1))|^2
\ee
in this case.

The simplest class of twisted lenses is $L^3_p(2)$, $p$ odd. Here,
$G=g_{\alpha_1\alpha_3}\otimes g_{\alpha_2\alpha_4}$, with the defining
relation

\be
g_{\alpha_1\alpha_3}\otimes g_{\alpha_2\alpha_4}\otimes
g_{\alpha_3\alpha_5}\otimes \ldots\otimes
g_{\alpha_{p-1}\alpha'_1}\otimes g_{\alpha_p\alpha'_2}
=\delta_{\alpha_1\alpha'_1}\delta_{\alpha_2\alpha'_2}
\ee
The corresponding link is shown in Fig.~4. One component is
unframed, the other has the framing $p$. In this case, it is tedious to
check the factorisation of links directly as it was done for
$L^3_p(1)$'s by applying the Kirby moves.

\nopagebreak
One can proceed for other lenses by analogy.

\newpage
\stepcounter{chapter}

\vspace*{4pc}

{\Large\bf Generating function for simplicial complexes.}

\vspace*{2pc}

Following ref. \cite{B} we can define the zero-dimensional field model
generating all possible
simplicial complexes weighted with the partition function (\ref{Z0}). Let
$\phi(x,y,z)$ be a ``translational'' invariant function on a quantum
group $G$, {\em i.e.}

\be
\phi(x,y,z)=\phi(xu,yu,zu)\hspace{1cm}\forall x,y,z,u\in G
\label{shiftinvar}
\ee
It can be represented in terms of matrix elements as follows.

\[
\phi(x,y,z)=\sum_{ \mbox{}^{\R_1\R_2\R_3}_{\{a_i,b_i\}}}
\sqrt{d_{R_1}d_{R_2}}
\varphi^{\R_1\R_2\R_3}_{a_1a_2a_3}
D^{R_1}_{a_1b_1}(x)D^{R_2}_{a_2b_2}(y)D^{R_3}_{a_3b_3}(z)
\overline{\langle b_1R_1b_2R_2|\overline{b}_3R_3\rangle}
\]\be
\langle \overline{b}_3R_3b_3R_3|0\rangle
\rule[-1.5cm]{0cm}{3cm}
= \sum_{R_1R_2R_3}\sqrt{\frac{d_{R_1}d_{R_2}}{d_{R_3}}}\
\fbox{$x$}\ \fbox{$y$}\ \fbox{$z$}\
\label{phi}
\ee
Eq. (\ref{phi}) is a general Fourier decomposition of a function
obeying eq. (\ref{shiftinvar}). We used for Fourier coefficients a
representation similar to the one for 3-$j$ symbols

\be
\rule[-1.5cm]{0cm}{3cm}
\varphi^{\R_1\R_2\R_3}_{a_1a_2a_3}=\frac{1}{\sqrt{d_{R_3}}}\hspace{3cm}
\overline{\varphi}^{\R_1\R_2\R_3}_{a_1a_2a_3}=
\frac{1}{d_{R_3}}\hspace{3cm}
\ee

We also demand  $\phi(x,y,z)$ to be symmetric with respect to even
permutations of arguments

\be
\phi(x,y,z)=\phi(z,x,y)=\phi(y,z,x)
\label{cycsym}
\ee

The first equality can be represented graphically as follows

\be
\rule[-1.5cm]{0cm}{3cm}
\frac{1}{d_{R_3}}\ \ \fbox{$x$}\ \ \fbox{$y$}\ \ \fbox{$z$}\
=\frac{1}{d_{R_3}}\ \ \fbox{$z$}\ \ \fbox{$x$}\ \ \fbox{$y$}\
=\frac{1}{d_{R_2}}\ \ \fbox{$z$}\ \ \fbox{$x$}\ \ \fbox{$y$}\
\ee
An odd permutation is equivalent to the complex conjugation:

\be
\overline{\phi}(x,y,z)=\phi(y,x,z)
\label{conj}
\ee
or, graphically,

\be
\rule[-1.5cm]{0cm}{3cm}
\frac{1}{d_{R_3}}\ \ \fbox{$x$}\ \ \fbox{$y$}\ \ \fbox{$z$}\
=\frac{1}{d_{R_3}}\ \fbox{$x^+$}\ \fbox{$y^+$}\ \fbox{$z^+$}\
=\frac{1}{d_{R_3}}\ \ \fbox{$y$}\ \ \fbox{$x$}\ \ \fbox{$z$}\
\ee
Eqs. (\ref{cycsym}) and (\ref{conj}) mean that the Fourier coefficients
have all symmetries of 3-$j$ symbols but their lower indices are
unrestricted.

We define the partition function by the integral

\be
P=\int {\cal D}\phi e^{-S}
\label{parfun}
\ee
where the action is taken in the form

\[
S=\frac{1}{2}\int dxdydz\; |\phi(x,y,z)|^2-\hspace{4cm}
\]\be
-\frac{\lambda}{12}
\int dxdydzdudvdw\; \phi(x,y,z)\phi(x,u,v)\phi(y,v,w)\phi(z,w,u)
\label{action}
\ee
The first term in eq. (\ref{action}) can be interpreted as two glued
triangles and the second, as four triangles forming a tetrahedron. It is
not surprising that, after the Fourier transformation, we get a 6-$j$
symbol attached to it:

\be
\rule[-2cm]{0cm}{4cm}
S=\sum_{ R_1R_2R_3}\hspace{2cm}-\frac{\lambda}{6}
\sum_{R_1\ldots R_6}\frac{1}{d_{R_1}^2d_{R_2}d_{R_3}}
\hspace{5cm}
\label{graphact}
\ee
The measure can be written in terms of Fourier coefficients

\be
{\cal D}\phi =\prod_{\mbox{}^{\R_1\R_2\R_3}_{a_1a_2a_3}}
d\varphi^{\R_1\R_2\R_3}_{a_1a_2a_3}
\label{measure}
\ee
If $q^n=1$, the product in eq. (\ref{measure}) runs over irreps from
the ground ring and, hence, is finite.

Practically, the partition function (\ref{parfun}) can be defined
within the perturbation expansion in $\lambda$. Two point correlator is
given by

\[
\rule[-1.5cm]{0cm}{3cm}
\frac{1}{P}\int {\cal D}\phi\;
\overline{\varphi}^{\R_1\R_2\R_3}_{a_1a_2a_3}
\varphi^{\scs \T_1\T_2\T_3}_{b_1b_2b_3}\;e^{-S}=\frac{1}{d_{R_3}}
\Bigg\langle\hspace{5cm}\Bigg\rangle=
\]\be
\rule[-1.5cm]{0cm}{3cm}
=\delta_{\R_1,\T_1}\delta_{\R_2,\T_2}\delta_{\R_3,\T_3}
\hspace{4cm}
\label{corr}
\ee
Performing all possible Wick pairings with the correlator (\ref{corr}),
we get in every order in $\lambda$ all oriended (due to eq.
(\ref{conj})) simplicial complexes.
For every bond in a simplicial complex,
we have a loop carrying a representation index.
It gives us a corresponding quantum dimension.
We have already had a 6-$j$ symbol inside a tetrahedron in
eq. (\ref{graphact}). Summing over all representations at bonds,
we reproduce the Turaev-Viro partition
function for a given simplicial complex. However, occasionally, we can
obtain non-trivial framings of the index loops, the ingredient absent in
the original Turaev-Viro partition function. The change of a framing
of a loop carrying an index $R$ by $\pm 1$ gives the factor
$q^{\pm C_R}$ ($C_R$ is a second Casimir). We can make all framings
trivial getting some factor, which can be as well attached to loops
appearing in the gauge theory framework of previous sections. We call
the generalized Turaev-Viro invariant the construction taking framings
into account.

Most of complexes are not manifolds but we do not need to improve the
technique developed in previous sections. By construction, all possible
simplicial complexes obey the restrictions imposed in Section 2 on
lattices (which are dual to the complexes). Gauge variables are attached
now to triangles and their relative order around bonds is always
unambiguously defined. We see that $\log P$ is the generating function
of 3-$d$ simplicial complexes weighted with the Turaev-Viro invariant.

It can be easily seen that simplicial complexes have non-negative Euler
character

\be
\chi=\sum_{i=1}^{N_0} p_i \geq 0
\label{euler}
\ee
where the sum runs over all vertices (0-simplexes) in a complex ($N_0$ is
their number). All tetrahedra touching an $i$-th vertex form a
3-$d$ ball; $p_i$ is the Euler character of its 2-$d$ boundary.
By definition, a complex is a manifold, if the vicinity of every
point is a sphere , {\em i.e.} $ p_i=0,\ \forall i$.

We see that non-manifolds can be obtained by inserting some local
topological defects or "vortices". In principle, such defects should
be given some energy depending on "the elasticity of the space". If we
consider $p_i$'s in eq. (\ref{euler}) as "vortex charges", it is natural
to estimate the non-manifolds contribution to the partition function as
if they were "vortex-like" particles:

\be
{\cal Z}(M^3)\approx \prod_{i=1}^{N_0}\sum_{p_i=0}^{\infty}
e^{-\varepsilon p_i}=\frac{1}{(1-e^{-\varepsilon})^{N_0}}
\label{vort}
\ee
where $M^3$ is some basic manifold. Of course, in eq. (\ref{vort}) there
should be subleading terms which can influence critical behavior. But we
see that non-manifolds cannot "blow up" our simplicial world.
For our model, $\varepsilon$ is an extra parameter. As was shown in
ref. \cite{B}, the role of $e^{-\varepsilon}$ can be partially played by
the rank of a gauge group, $|G|$.

Unfortunately, the partition function (\ref{parfun}) hardly can be found
in a closed form. However, this model gives a framework for the strong
coupling expansion in simplicial gravity.

\vspace*{4pc}

{\Large\bf Concluding remarks}
\nopagebreak
\vspace*{2pc}

In this paper, we consider only closed oriented manifolds. One can
easily generalize our approach to manifolds with boundaries. If we fixe
a field configuration on boundaries,
the partition function (\ref{Z0}) will be a function of corresponding
gauge variables. If a manifold is a ball with a connected boundary, then
we obtain a lattice regularization of the 2-dimensional WZNW
model. As our construction is based on the Turaev-Viro invariant, we
have both holomorphic and anti-holomorphic sectors of the model. If a
boundary has many connected components, one can study amplitudes. We
hope that our approach gives a covenient technique for practical
calculations.

The complete set of observables in gauge theory is given by correlators
of Wilson loop operators. The strightforward generalization to
the quantum case gives well known knot invariants in a manifold. On a
lattice, we can construct a Wilson loop operator simply multiplying
matrix elements along a loop. A framing is defined by a way
of the multiplication. The using of ribbon
graphs allow for its visualization.

Our intention in this paper was to develop a suitable framework for
further investigations  of intriguing connection between gauge
theories, 3-manifold invariants and quantum gravity.
We hope that, at least partially, we have done it.

\vspace*{4pc}

{\Large\bf Acknowledgements}
\nopagebreak
\vspace*{2pc}

Discussions with S.Piunikhin are greatly appreciated.

\vspace*{4pc}

\appendix
\stepcounter{chapter}
{\Large\bf Appendix}
\nopagebreak
\vspace*{2pc}

For reader's convenience, we collect in this appendix main formulas
which the aproach advocated in this paper gives in 2-dimensions.

The invariance under right shifts
\be
\phi(xu,yu)=\phi(x,y)
\ee
means here that we consider functions of one argument:
$\phi(x,y)=\phi(xy^+)$. The action in the model analogous to
(\ref{action}) takes the form

\be
S=\frac{1}{2}\int dx\; |\phi(x)|^2-\frac{\lambda}{3}\int dxdydz\;
\phi(xy^+)\phi(yz^+)\phi(zx^+)
\label{action2}
\ee
To get oriented surfaces, we demand

\be
\phi(x^+)=\overline{\phi}(x)
\label{conj1}
\ee
then, if we adopt the following graphical represenation,

\be
\rule[-1.5cm]{0cm}{3cm}
\phi(x)=\sum_{R;ab\in V_R}\sqrt{d_R} \varphi^\R_{ab} D^R_{ab}(x)=
\sum_{R;ab\in V_R}\sqrt{d_R}\ \fbox{$x$}\hspace{3cm}
\ee
eq. (\ref{conj1}) gives the hermiticity condition for Fourier
coefficients:
\be
\rule[-1.5cm]{0cm}{3cm}
\fbox{$x$}\hspace{2cm}=\ \fbox{$x^+$}\hspace{2cm}=\ \fbox{$x$}\hspace{2cm}
\ee

If $q=e^{i\frac{2\pi}{n}}$, the Fourier transform of eq.
(\ref{action2}) is just a number of copies of the $\phi^3$
matrix model action with different matrix sizes given by dimensions of
representation spaces, $dim(V_R)$:

\be
\rule[-1.5cm]{0cm}{3cm}
S=\sum_R\Bigg\{\frac{1}{2}\hspace{3cm} -\frac{\lambda}{3\sqrt{d_R}}
\hspace{4cm}\Bigg\}
\ee
The partition function is still defined by eq. (\ref{parfun}).
After expanding in $\lambda$, we get an index loop around each vertex of
a triangulation. It gives the quantum dimension, $d_R$, of a
representation, $R$,
carried by the loop. In other respects, the model is quite analogous to the
ordinary matrix model, and we find its free energy to be simply

\be
\log P=\sum_R \log P_{d_R}
\ee
where $P_N$ is the $N\times N$-matrix model partition function continued
analytically from integer sizes of matrices, $N$, onto the whole complex
plane.

Surfaces generated by this model are weighted with the
``quantum'' invariant:

\be
{\cal I}(M^2_p)=\int_G \prod_{k=1}^p du_p dv_p\; \delta(\prod_{k=1}^p
u_kv_ku^{-1}_kv^{-1}_k, I)=\sum_R (d_R)^{2(1-p)}
\label{2invar}
\ee
where we use the standard cell decomposition of an oriented surface,
$M^2_p$, with $p$ handles. It consists of one 0-cell, one 2-cell and $2p$
1-cells: $\sigma^1_{u_k},\ \sigma^1_{v_k};\ k=1,\ldots,p$.
The only defining relation is coded in the agrument of the
$\delta$-function. The result depends only on the Euler character,
$\chi=2(1-p)$, of $M^2_p$.

Calculating a \mbox{2-dimensional} gauge partition function,
we actually repeate the
corresponding calculations with classical groups \cite{Mig}, because all
appearing links are trivial.
However, in our case, the number of representations is finite, and eq.
(\ref{2invar}) does not need to be regularized. Hence, we have well
defined \mbox{2-dimensional} topological gauge theory.

\newpage
\vspace*{3pc}
\begin{center}
{\Large\bf Figure Captions}
\end{center}
\vspace{3pc}
\begin{enumerate}
\item An example of the shift of variables (\ref{varout}).
Ribbons are used to show framings.\\
\bigskip
\item An example of the division of a 2-cell before its shrinking.\\
\bigskip
\item A link representation of ${\cal I}(L^3_p(1))$. The dashed lines
show wires to perform the $\beta$-move disjoining the loops.\\
\bigskip
\item A link representing $L^3_p(2)$.\\
\end{enumerate}
\end{document}